\documentstyle [12pt]{article}

\begin{document}

\baselineskip 7.5 mm

\def\thefootnote{\fnsymbol{footnote}}

\begin{flushright}
\begin{tabular}{l}
CERN-TH/97-346 \\
hep-ph/9712212
\end{tabular}
\end{flushright}

\vspace{12mm}

\begin{center}

{\Large \bf

Experimental signatures of supersymmetric dark-matter Q-balls 
}
\vspace{18mm}

\setcounter{footnote}{0}

{\large
Alexander Kusenko,}$^1$\footnote{ email address: Alexander.Kusenko@cern.ch}
{\large
Vadim Kuzmin,}$^2$\footnote{ email address: kuzmin@ms2.inr.ac.ru } 
{\large
Mikhail Shaposhnikov,}$^{1}$\footnote{ email address:
  mshaposh@nxth04.cern.ch} \\
{\large and P.~G.~Tinyakov}$^{2}$\footnote{email address: 
peter@flint.inr.ac.ru}

\vspace{4mm}
$^1$Theory Division, CERN, CH-1211 Geneva 23, Switzerland \\
$^2$Institute for Nuclear Research, 60th October Anniversary Prospect 7a,
Moscow, Russia 117312

\vspace{26mm}

{\bf Abstract}
\end{center}

Theories with low-energy supersymmetry predict the existence of stable 
non-topological solitons, Q-balls, that can contribute to dark matter. 
We discuss the experimental signatures, methods of detection, and the
present limits on such dark matter candidates.

\vspace{20mm}

\begin{flushleft}
\begin{tabular}{l}
CERN-TH/97-346 \\
November, 1997
\end{tabular}
\end{flushleft}

\vfill

\pagestyle{empty}

\pagebreak

\pagestyle{plain}
\pagenumbering{arabic}
\renewcommand{\thefootnote}{\arabic{footnote}}
\setcounter{footnote}{0}

Supersymmetric generalizations of the Standard Model, in particular 
the minimal version, MSSM, invariably predict the existence of
non-topological solitons, dubbed Q-balls~\cite{coleman}, with an arbitrary
baryon number~\cite{ak_mssm}.   Supersymmetric Q-balls are coherent states
of squarks, sleptons, and the Higgs fields.  In theories with ``flat
directions'' in the scalar potential, which are generic for supersymmetry,
these objects may exhibit a number of interesting properties
\cite{dks,ak_other}.  In particular, solitons with a large baryon number 
are entirely stable~\cite{ks} and can be copiously produced in the early
universe~\cite{ks}.  This makes relic Q-balls an appealing candidate for
cold dark matter~\cite{ks}.  In this Letter we will examine the
implications of this speculative type of dark matter  for detector
experiments.   

Flat potentials $U(\phi)$, {\it i.\,e.}, those that grow slower than the
second power of the scalar VEV $\phi$, arise naturally in theories with
low-energy supersymmetry breaking (see, {\it e.\,g.}, Refs.~\cite{gia1,gmm}
and discussion in  Ref.~\cite{dks}).  
For example, if $U(\phi) \sim m^4=const$ for large $\phi$,  
the mass of a soliton with charge (baryon number) $Q_{_B}$ is 
$M_{_Q}  \simeq (4\pi \sqrt{2}/3) \, m \, Q_{_B}^{3/4}$, its radius is 
$R_{_Q}  \simeq  (1/\sqrt{2})\,  m^{-1} \,  Q_{_B}^{1/4}$, and the
maximal scalar VEV inside is $\phi_{_Q}  \simeq (1/\sqrt{2}) \,  m \,
Q_{_B}^{1/4}$.  We will assume these
relations and neglect the logarithmic corrections to the flat potentials
that appear in realistic theories~\cite{gia1,gmm}.  One assumes $m$ to be 
from $100$~GeV to $100$~TeV, higher values being disfavored by the
naturalness arguments. For a specific model of supersymmetry breaking
studied in Ref.~\cite{gmm}, $m \sim 1$~TeV.  

The baryon number $Q_{_B}$ of a stable soliton must be
greater than $10^{15} (m/1 {\rm TeV})^4$~\cite{ks}.  Larger solitons cannot
decay into the matter fermions because the energy per unit baryon number is
less than the proton mass.  Q-balls with a much greater global charge, in
excess of  $10^{20}$, can be produced in the early universe from the
breakdown of a coherent scalar condensate~\cite{ks}.  Formation of such
condensate,  being the starting point of the Affleck--Dine scenario for
baryogenesis \cite{ad}, may also explain the baryon asymmetry of the
universe, in which case the initial baryon number stored in the condensate
is distributed between the matter baryons and Q-balls.  If the ordinary
baryonic matter and the dark matter share the same origin~\cite{ks}, one
may hope to  explain why the two have, roughly, the same density in the
Universe.

The flux of cosmic Q-balls falling on Earth can be estimated under the
assumption that they make a sizeable contribution to the missing matter of
the universe.  As follows from Ref.~\cite{ks}, Q-balls produced from the
breakdown of a primordial condensate have a very narrow distribution of
charges.  We will assume, therefore, that all dark-matter solitons have the
same mass.  Q-balls can be of interest as dark matter
candidates if their mass density in the galactic halo is of order
$\rho_{_{DM}} \approx 0.3$~GeV/cm$^3$, which corresponds to the number
density  

\begin{equation}
n_{_Q}\sim \frac{\rho_{_{DM}}}{M_{_Q}} \sim 5 \times 10^{-5} \, Q_{_B}^{-3/4} 
\left ( \frac{1 {\rm TeV}}{m}
\right ) {\rm cm}^{-3}.
\label{num_dens}
\end{equation}
We assume the average velocity for Q-balls $v \sim 10^{-3} \, c$.  Then 
the flux is 
$F\simeq (1/4\pi) n_{_Q} v \sim  10^{2} \, Q_{_B}^{-3/4} \left ( \frac{1 {\rm
      TeV}}{m} \right ) {\rm cm}^{-2} {\rm s}^{-1} {\rm sr}^{-1}$.
For example, the total surface area of the water tank used in the
Super-Kamiokande experiment~\cite{sk} is $7.5 \times 10^7$ cm$^2$.  If all or
most of the dark matter is made up of solitons with charge $Q_{_B}$, some
Q-balls must go through this detector at the rate  

\begin{equation}
N\sim \left ( \frac{10^{24}}{Q_{_B}}  
\right )^{3/4} \left ( \frac{1 {\rm TeV}}{m} \right ) {\rm yr}^{-1} .
\label{events}
\end{equation}
Q-balls can also produce a signal, at a comparable rate, at the Baikal Deep
Underwater Neutrino Experiment~\cite{baikal}, as well as other experiments.  

Let us consider the interactions of baryonic solitons with ordinary matter. 
The interior of a large Q-ball can be thought of as a spherically-symmetric
region filled with a non-standard vacuum that breaks spontaneously the 
baryonic $U(1)_{_B}$ symmetry.  The scalar VEV inside a stable soliton 
extends along a flat direction in the MSSM scalar potential 
and carries the corresponding quantum numbers.  

If supersymmetry is exact (which we assume to be the case
for sufficiently large VEV, as in theories with SUSY breaking communicated
at low energy), the MSSM has a very large space of degenerate 
vacua, the flat directions, labelled by the corresponding  gauge-invariant
holomorphic polynomials of the chiral superfields~\cite{flat1,flat2}. 
They have been enumerated and catalogued in
Ref.~\cite{drt,flat_mssm}.  Each flat direction is parameterized by a
gauge-invariant scalar VEV. Those that carry some baryon number can give
rise to stable Q-balls, and may also play a central role in generating
baryon asymmetry of the Universe~\cite{ad,gmm,drt}.  

Inside a Q-ball the SU(3)$\times$SU(2)$\times$U(1) 
gauge symmetry may be broken by the VEV of squarks, sleptons, and the Higgs
fields. In the absence of fundamental SU(3)-singlet baryons in the MSSM, 
any baryonic Q-ball has a broken SU(3) inside.  In contrast, the
electroweak symmetry may be restored if the only fields that have non-zero
VEV are SU(2) singlets.  This is the case for Q-balls that
have a scalar VEV aligned, for example, with the $udd$ flat direction 
(notation of Ref.~\cite{flat_mssm}).  Although baryon number is violated by
the instantons, the rate is suppressed because the the size of the
instantons that fit inside a Q-ball is small.  

A baryonic Q-ball must have a non-zero VEV $\phi_{_Q} \sim m \, Q_{_B}^{1/4}$ 
of scalar quarks in its interior.  It may or may not be accompanied by the
VEV's of sleptons and the Higgs fields.  Matter fermions cannot penetrate
inside some Q-balls because their masses inside may be increased by the
large Higgs VEV, as well as through their mixing with gauge fermions.  
However, the outer region of any Q-ball  has a layer near its
boundary where (i) the quark masses are less than $\Lambda_{_{QCD}}$ and 
(ii) the gauge SU(3) symmetry is broken spontaneously by the VEV's of
squarks.  When a nucleon enters this region, where the QCD deconfinement
takes place, it dissociates into quarks.  The energy released in such
process, roughly 1 GeV per nucleon, is emitted in pions.  This process is
the basis for the experimental detection of the dark-matter Q-balls. 

As an electrically neutral Q-ball passes through matter, it absorbs the
nuclei with a cross-section determined entirely by the soliton's size,
$\sigma \sim 10^{-33} \, Q_{_B}^{1/2} (1 {\rm TeV}/m)^2$~cm$^{2}$.  The
corresponding mean free path in matter with density $\rho $ is  

\begin{equation}
\lambda_0 \sim 10^{-3} \, A \, \left ( \frac{10^{24}}{Q_{_B}}
\right )^{1/2}  
\left ( \frac{m}{ 1 \ {\rm TeV}}
\right )^2 
\left ( \frac{1 \ {\rm g}/{\rm cm}^3}{\rho}
\right ) \ {\rm cm}, 
\label{lambda}
\end{equation}
where $A$ is the weight of the nucleus in atomic units. 
The quarks caught in the deconfining coat of a Q-ball are absorbed into the
condensate eventually via the reaction $qq \rightarrow \tilde{q} 
\tilde{q}$ that proceeds with a (heavy) gluino exchange.  The reason this
process is energetically allowed is, of course, because the squarks in the
condensate are nearly massless. The rate of conversion is suppressed by the
square of the gluino mass. If the condensate
in the Q-ball is different in flavor from the quarks, an additional 
CKM suppression takes place.  In any case, the absorption of quarks
into the condensate occurs at a much higher rate than the collisions 
of Q-balls with nuclei characterized by $\lambda_0$ in equation 
(\ref{lambda}).  

For energetic reasons, large Q-balls comprise an electrically neutral scalar
condensate.  However, unless the electrons are trapped by the Q-ball, 
the process described above proceeds through the formation of a bound state
of the Q-ball to quarks which has a positive electric charge. 
If this is the case, the electrons can be captured eventually in an
electroweak process $u e \rightarrow d \nu$ which, we note in passing, 
is very fast inside those Q-balls that restore the SU(2) gauge symmetry 
because the $W$ boson is massless. 
 
However, the electrons cannot penetrate inside those
Q-balls, whose scalar VEV gives them a large mass.  For example, the 
simultaneously large VEV's of both the left-handed ($L_e$) and the
right-handed ($e$) selectrons along the $QQQLLLe$ flat direction give rise
to a large electron mass through mixing with the gauginos.  The  locked out
electrons can form bound states in the Coulomb field of the (now
electrically charged) soliton.  The resulting system is similar to an atom
with an enormously heavy nucleus. Based on their ability to retain  
electric charge, the relic solitons can be separated in two classes:
Supersymmetric Electrically Neutral Solitons  (SENS) and Supersymmetric
Electrically Charged Solitons (SECS).  The interactions of Q-balls with
matter,  and, hence, the modes of their  detection, differ drastically
depending on whether the dark matter comprises SENS or SECS.   

First, the Coulomb barrier can prevent the absorption of the 
incoming nuclei by SECS.  A Q-ball with baryon number $Q_{_B}$ and
electric charge $Z_{_Q}$ cannot imbibe protons moving with velocity $v \sim
10^{-3} c$  if $Q_{_B} \stackrel{<}{_{\scriptstyle \sim}} 10^{29} Z_{_Q}^4
(m/1\, {\rm TeV})^4$.  Second, the scattering cross-section of an
electrically charged Q-ball passing through matter is now determined by,
roughly, the Bohr's  radius, rather than the Q-ball size: $\sigma \sim \pi
r_{_B}^2 \sim 10^{-16} cm^2$.  The corresponding mean free path is 

\begin{equation}
\lambda_e \sim 10^{-8} A 
\left ( \frac{1 \ {\rm g}/{\rm cm}^3}{\rho}
\right ) \ {\rm cm}. 
\label{lambda_e}
\end{equation}
By numerical coincidence, the total energy released 
per unit length of the track in the medium of density $\rho$ is, roughly,
the same for SENS and SECS,
$dE/dl \sim $ $100 \, (\rho/1 \, {\rm g\,  cm}^{-3}) \,$ ${\rm GeV/cm}$. 
However, the former takes in nuclei and emits pions, while the latter 
dissipates its energy in collisions with the matter atoms. 
Signatures of baryonic and anti-baryonic solitons are expected to be 
similar. 

A passage of a Q-ball with baryon number $Q_{_B}\sim 10^{24}$ through a
detector,  associated with emission of, roughly, 10 GeV per millimeter 
can make a spectacular signature.  Of course, depending on the mass
parameter $m$ and the charge $Q_{_B}$, the frequency of such events can be
small; for some values, too small to be detected.  
As is evident from equation (\ref{events})  
the generic values of parameters are not ruled out, and  are consistent
with observation of relic Q-balls at the existing and future facilities.
Since the anticipated tracks are very energetic and unmistakable, it is the
surface area of the detector, rather than its fiducial volume, that is
important.  A large-area detector (LAD) would, in general, be more
effective in searching for Q-balls than a more compact machine with the
same volume.   

The present experimental limit on the flux of SECS is set by
the MACRO search~\cite{macro} for ``nuclearites''~\cite{dg}, which have
similar interactions with matter: 
$F < 1.1 \times 10^{-14}$ cm$^{-2}$~s$^{-1}$~sr$^{-1}$.  This translates
into the lower limit on the baryon number of dark-matter Q-balls, $Q_{_B} 
\stackrel{>}{_{\scriptstyle \sim}} 10^{21}$.  Signatures of SENS are 
similar to those expected from the Grand Unified monopoles that catalyze the
proton decay.  If one translates the current experimental limits from
Baikal~\cite{baikal} on the monopole flux, one can set a limit on the charge 
of SENS, $Q_{_B} \stackrel{>}{_{\scriptstyle \sim}} 3 \times 10^{22}$, for
$m=1~{\rm TeV}$.  Non-observation 
of Q-balls at the Super-Kamiokande after a year of running would improve
this limit by two orders of magnitude.   Of course, this does not
preclude the existence of smaller Q-balls with lower abundances that 
give negligible contribution to the matter density of the universe. 

Electrically charged Q-balls with a smaller baryon number can dissipate
energy so efficiently  that they may never reach the detector. 
SECS with baryon number $Q_{_B} \stackrel{<}{_{\scriptstyle \sim}}
10^{13} (m/1\, {\rm TeV})^{-4/3}$ can be stopped by the 1000 m of
water equivalent matter shielding.  Such solitons could not have been
observed by the underground detectors.  Therefore, in the window of
$Q_{_B}\sim  10^{12}...10^{13}$ the flux of SECS appears to be virtually
unconstrained. 

For completeness, we will briefly review some astrophysical constraints. 
A SENS that passes through Earth with velocity $10^{-3} \, c$  
looses a negligible part of its kinetic energy to collisions with the matter
particles.  The total change in its velocity is $\delta v/v \sim 10^{-2} 
Q_{_B}^{-1/4} (1 \, {\rm TeV}/m)^3$.  Therefore, SENS do not
accumulate inside ordinary stars and planets.  A neutron star is
sufficiently dense to stop a Q-ball of any kind.  During the 
period of $10^{8}$ years (the age of the oldest observed pulsars) of order
$\sim 10^{33} Q_{_B}^{-3/4} (1\, {\rm TeV}/m)$  relic solitons are captured
by a neutron star.  Since the nuclear matter is very dense, the energy
released in the capture of nucleons by the Q-balls is significantly higher
than that in the ordinary matter.  The interactions of the relic Q-balls
with neutron stars and white dwarfs are studied in Ref.~\cite{starwreck}.

However, the combined heat from all
Q-balls captured in 100 Myr can lead to an increase in temperature of the
neutron star by only $\stackrel{<}{_{\scriptstyle \sim}} 0.01 
(Q_{_B}/10^{24})^{-1/16} $ keV, too small to have any observable
consequences.   

SECS's do accumulate in ordinary stars.  However, the Coulomb barrier
prevents a rapid absorption of nuclei and inhibits the production of pions.
Therefore, in contrast to the case of monopoles, there is no constraint on
the abundance of SECS from observations of the low-energy solar neutrinos. 

It should also be mentioned that, because of its very large mass, a Q-ball
passing through the atmosphere cannot create an extensive shower typical
for the high-energy cosmic rays.  The effectiveness of the wide-array
detectors in searching for Q-balls is, therefore, limited by the total area
of their counters.  Searches for stable ultra-heavy nuclei in matter
\cite{heavy}, which may be suitable for detecting smaller Q-balls (with
charges $10^{12}...10^{13}$), 
afford no limit at present because the mass range of
interest, $m_{_Q} \stackrel{>}{_{\scriptstyle \sim}} 10^{12}$~GeV, 
has never been explored.  

It would be interesting to see if some of the exotic events in the cosmic
rays, {\it e.\,g.}, the so called Centauro events~\cite{centauro}, 
the penetrating halo event of the Pamir experiment~\cite{pamir,centauro}, 
and the ultra-high energy cosmic rays that appear to defy the GZK 
bound~\cite{gzk}, may be related to the relic Q-balls.  

In summary, Q-ball is an appealing dark matter candidate predicted by
supersymmetry.  Baryonic Q-balls have strong interactions with matter and 
can be detected in present or future experiments.  
Observational signatures of the baryonic solitons are characterized by a
substantial energy release along a straight track with no attenuation
throughout the detector.  The present experimental
lower bound on the baryon number $Q_{_B} \stackrel{>}{_{\scriptstyle \sim}}
10^{21}$ is consistent  with theoretical expectations~\cite{ks} for the
cosmologically interesting range of Q-balls in dark matter.   
In addition, smaller Q-balls, with the abundances much lower than that in 
equation (\ref{num_dens}), can be present in the universe.  Although
their contribution to $\Omega_{_{DM}}$ is negligible, 
their detection could help unveil the history 
of the universe in the early post-inflationary epoch.  Since the breakdown
of a coherent scalar condensate~\cite{ks} is the only conceivable mechanism
that could lead to the formation of Q-balls with large global charges, 
the observation of any Q-balls would seem to speak unambiguously 
in favor of such process having taken place.  This would, in turn, have
far-reaching implications for understanding the origin of the baryon
asymmetry of the universe, for the theory of inflation, and for 
cosmology in general.

We thank W.~Frati, T.~Gherghetta, J.~Hill, A.~Smirnov, and E.~Witten for
discussions.  V.~K. and P.~G.~T. thank Theory Division at CERN for
hospitality.

\end{document}